# IDENTIFICACIÓN DE CELDAS SEMILLA EN IMÁGENES MULTIESPECTRALES PARA LA SEGMENTACIÓN CON GROWCUT

*Wuilian Torres[1], Antonio Rueda-Toicen[2,3]*

[1]Centro de Procesamiento Digital de Imágenes. Instituto de Ingeniería. e-mail: wuiliantor@gmail.com

[2]Instituto Nacional de Bioingeniería, Universidad Central de Venezuela.

[3]Algorithmic Nature Lab, LABORES for the Natural and Digital Sciences, Paris, France

e-mail: antonio.rueda.toicen@algorithmicnaturelab.org

## RESUMEN

La segmentación de las imágenes satelitales es el paso previo para la clasificación de imágenes orientada a objetos. Está técnica ha cobrado auge durante los últimos años por su aplicabilidad al procesamiento de imágenes con alta resolución espacial. Para su aplicación la imagen debe ser segmentada en regiones uniformes, actividad que requiere alta interacción de parte del usuario, quien debe explorar exhaustivamente la imagen para establecer de manera interactiva umbrales que permitan obtener imágenes procesadas que no estén sobre-segmentadas y que no obvien segmentos representativos para el estudio. En este trabajo se propone una técnica que tiene por objetivo la segmentación automática de la imagen multiespectral; se inicia con la identificación en la imagen de zonas homogéneas en cuanto a su firma espectral, utilizando para ello filtros morfológicos. Estas zonas homogéneas son representantes de los diferentes tipos de cobertura presentes en la imagen y se utilizan como semillas en un algoritmo de segmentación multiespectral denominado GrowCut. El GrowCut es un autómata celular con crecimiento competitivo de regiones, sus celdas están asociadas a cada píxel de la imagen mediante tres parámetros: la firma espectral, la etiqueta y un factor que indica la fortaleza con la cual la celda recibe una etiqueta. Las celdas semilla poseen la máxima fortaleza y mantienen su estado durante toda la evolución del autómata. Partiendo de las celdas semillas, de manera iterativa cada celda de la imagen es atacada por las celdas vecinas, cuando su fortaleza es inferior a la fuerza de ataque de alguna de sus vecinas cambia su estado asumiendo la etiqueta del más fuerte y tomando la fortaleza con la que fue vencida. La función que permite establecer la fuerza del ataque es mayor mientras menor es la distancia espectral entre los píxeles. Una vez que el autómata llega a su equilibrio se obtiene la imagen segmentada donde cada píxel ha tomado la etiqueta de alguna de las semillas. En este trabajo se aplicó el algoritmo en una imagen adquirida por el Landsat 8 en la región agrícola de Calabozo en el estado Guárico, Venezuela; se presentan diferentes tipos de coberturas: agricultura, centros poblados, cuerpos de agua y sabanas con diferentes grados de intervención. Se muestra a la segmentación obtenida como un conjunto de polígonos irregulares que encierran a los objetos geográficos.

Palabras clave: Segmentación, GrowCut, Filtros Morfológicos, Imagen Multiespectral, Autómata celular

## IDENTIFICATION OF SEED CELLS IN MULTISPECTRAL IMAGES FOR GROWCUT SEGMENTATION

### ABSTRACT

The segmentation of satellite images is a necessary step to perform object-oriented image classification, which has become relevant due to its applicability on images with a high spatial resolution. To perform object-oriented image classification, the studied image must first be segmented in uniform regions. This segmentation requires manual work by an expert user, who must exhaustively explore the image to establish thresholds that generate useful and representative segments without oversegmenting and without discarding representative segments. In this work, we propose a technique that automatically segments the multispectral image while facing these challenges. We start identifying in the image homogenous zone according to their spectral signatures through the use of morphological filters. These homogenous zones are representatives of the different types of land coverings in the image and are used as seeds for the GrowCut multispectral segmentation algorithm. GrowCut is a cellular automaton with competitive region growth, its cells are linked to every pixel in the image through three parameters: the spectral signature of the pixel, a label, and a strength factor that represents the strength with which a cell defends its label. The seed cells possess maximum strength and maintain their state throughout the automaton's evolution. Starting from seed cells, each cell in the image is iteratively attacked by its neighboring cells. When the defending cell has lower strength than the strength of attack, it takes the label of the attacking cell and updates its strength as equal to the strength of the defeating attack. The attack strength function returns values that are inversely proportional to the distance between pixels, meaning that pixels with similar spectral signatures are likely to take the same label. When the automaton stops updating its states, we obtain a segmented image where each pixel has taken the label of one of its cells. In this paper the algorithm was applied in an image acquired by Landsat8 on agricultural land of Calabozo, Guarico, Venezuela where there are different types of land coverings: agriculture, urban regions, water bodies, and savannas with different degrees of human intervention. The segmentation obtained is presented as irregular polygons enclosing geographical objects.

## INTRODUCCIÓN

La necesidad de disponer información detallada sobre lo que acontece en la superficie de la tierra a nivel local, global o regional, ha impulsado el desarrollo de sensores que proveen imágenes con mayor resolución espacial. Las técnicas clásicas para su análisis se han orientado al estudio de los parámetros estadísticos relacionados con la firma espectral de cada uno de sus píxeles sin tomar en cuenta el contexto espacial; este enfoque pierde efectividad cuando el tamaño del píxel supera ampliamente el mínimo requerido para cumplir con las exigencias de la escala de trabajo (Blaschke et al., 2010). Las técnicas de análisis de imagen orientadas a objetos: Object-Based Image Analysis: OBIA o Geographic OBIA: GEOBIA, incorporan métodos que permiten descomponer la imagen en zonas homogéneas disjuntas, denominadas objetos u objetos geográficos (OG). Se caracterizan por tener un tamaño acorde con la escala de trabajo y por englobar píxeles adyacentes con propiedades espectrales similares (Cartwright et al., 2008). Otros descriptores de los OG pueden ser incorporados en sus análisis, tales como la textura, tamaño, morfología, etc.  La primera fase del OBIA es crítica, la imagen es segmentada agrupando píxeles contiguos que poseen características similares, firma espectral, textura, estos segmentos se corresponden con los OG. En general, se requiere la intervención del usuario para evitar la sub y sobre-segmentación. En la literatura se dispone de una gran variedad de métodos para segmentar las imágenes, algunos siguen una estrategia orientadas a la identificación de las fronteras entre los objetos, otros se basan en la cuantificación de la similitud entre píxeles (He et al., 2014). Se dispone de algunos sistemas comerciales que permiten el análisis de imágenes orientado a objetos: eCognition, Imagine Objetive, ERDAS, y el módulo Feature extraction module de ENVI (Navulur, 2007), que en general requieren la asistencia del usuario para obtener una segmentación adecuada a sus requerimientos. Este trabajo es la continuación de la búsqueda de técnicas de segmentación no supervisadas con el fin de reducir los tiempos de análisis y aumentar la objetividad de los resultados, evitando la subjetividad introducida por el especialista en su interpretación (Torres, Rueda-Toicen, 2014). Se propone la utilización de autómatas celulares partiendo del algoritmo de GrowCut propuesto por Vezhnevets (Vezhnevets et al., 2005. El algoritmo propuesto analiza la imagen para identificar las zonas homogéneas representativas de cada OG presente en la imagen, las cuales constituirán las celdas o celdas semilla. Para ilustrar los resultados se utilizará la imagen adquirida el 31-01-2014 por el sensor OLI (Operational Land Imager) del Landsat 8; se extrajo una subimagen de dimensión 1800 x 2200 x 4 píxeles en Calabozo, estado Guárico, Venezuela, donde se extrajeron las bandas 2 a 5 de la imagen Landsat. La Figura 1a presenta la subimagen en falso color infrarrojo y la Figura 1b el detalle en una zona con alta diversidad de OG.

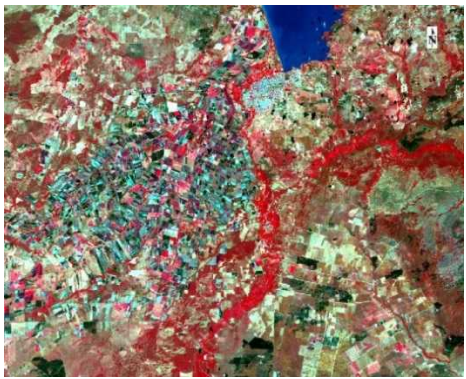
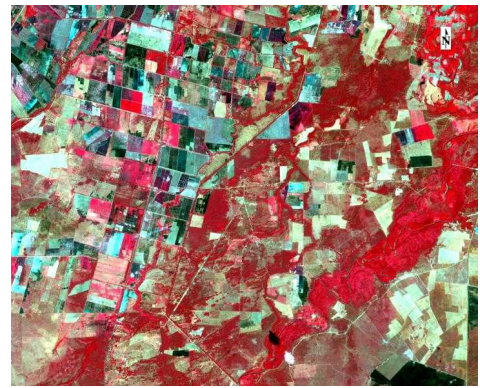

**Figura 1a** Sub-imagen de Calabozo, Edo. Guárico, Venezuela.   **1b.** Detalle de la imagen

## METODOLOGÍA

La metodología seguida para obtener las semillas y la segmentación de la imagen comprende tres fases: la detección de semillas que representaran cada una de los OG presentes en la imagen, el etiquetado de las semillas que tiene por objetivo disminuir la cantidad de segmentos agrupando las semillas en función de su firma espectral y la segmentación utilizando el autómata celular.

La imagen es filtrada mediante filtros morfológicos alternados secuenciales cuyas propiedades permiten mantener las fronteras entre los OG reduciendo las pequeñas variaciones de la reflectancia dentro de ellos. Las semillas son obtenidas a partir del gradiente morfológico de la imagen, corresponden a zonas homogéneas donde el gradiente toma valores cercanos a cero. Cada semilla representa un OG, las semillas con pocos píxeles son eliminadas para considerar solamente los OG de mayor superficie que están presentes en la imagen, reduciendo así la sobre-segmentación de la imagen.

Durante la fase de etiquetado se analizan las firmas espectrales de las semillas y se les asigna una etiqueta. Las semillas con firmas espectrales cercanas reciben la misma etiqueta.

A partir de las semillas se utiliza un autómata celular basado en el algoritmo de GrowCut para segmentar la imagen. Las celdas de cada semilla compiten entre ellas para tomar las celdas vecinas siguiendo una regla de evolución que considera la similitud de las firmas espectrales.

### El Autómata Celular GrowCut

En este trabajo la utilización de los autómatas celulares para la segmentación de imágenes multiespectrales tiene como principal objetivo el de reducir la subjetividad que introduce el especialista al definir los OG durante la interpretación visual de la imagen. Los autómatas celulares, introducidos originalmente por Von Neuman, (1966) se caracterizan por ser un retículo $n$-dimensional de celdas. Cada celda posee un estado discreto que es evaluado en un paso de evolución siguiendo reglas de actualización deterministas. Estas reglas determinan el estado de cada celda en el siguiente instante de evolución considerando el estado actual de la celda y otras celdas en su vecindad.

Vezhnevets y Konouchine, (2005) proponen un algoritmo para un autómata celular que permite separar, en una imagen $n$-dimensional, el objeto de interés siguiendo una regla de crecimiento competitivo de regiones denominada "GrowCut". Cada píxel de la imagen se asocia a una celda del autómata. En su estado inicial, se seleccionan manualmente celdas semillas representativas de las regiones presentes en la imagen. En cada iteración, para la evolución del autómata se evalúa la afinidad de las celdas semillas con sus vecinas para decidir si es anexada a su región. El autómata celular se define como una 3-tupla $A = (E, V, R)$, donde $E$ representa el conjunto de los estados del autómata en un instante $t$, $V$ la vecindad utilizada para evaluar cada celda y $R$ la función utilizada como regla de transición del estado de la celda al pasar del instante $t$ al $t+1$. El estado $E_p$ de la celda $p$ se define como una 3-tupla $(l_p, \theta_p, \vec{C}_p)$, $l_p$ corresponde a su etiqueta, $\theta_p$ es la fortaleza con la cual la celda fue asignada al estado actual con valores entre 0 y 1, y $\vec{C}_p$ es un vector que contiene los descriptores utilizados para analizar la celda (niveles digitales, índice de vegetación, textura, etc.). La imagen digital es una matriz cuyos $m \times n$ píxeles se asocian con un autómata que dispone de $P = m \times n$ celdas. El algoritmo de GrowCut que describe la regla para la evolución del autómata en el instante t es el siguiente:

$A$ = true
while ( $A$ )
    A = false
    // para cada celda p
for $\forall p \in P$
    // Copiar estado previo: etiqueta y fortaleza
    $l_p^{t+1} = l_p^t; \theta_p^{t+1} = \theta_p^t$
    // La celda vecina $q$ ataca a la celda $p$ con
    // una fuerza de ataque FA$_q$
    for $\forall q \in N(p)$
      if $FAq = g(|\vec{C}_p - \vec{C}_q|) \cdot \theta_q^t > \theta_p^t$
        $l_p^{t+1} = l_q^t$
        $\theta_p^{t+1} = g(|\vec{C}_p - \vec{C}_q|) \cdot \theta_q^t$
        A = true
      end if
    end for
  end for
end while

Donde $g$ es una función monótona decreciente con valores limitados entre 0 y 1 definida por la ecuación 1.

$$max \vee \overline{C} \vee ¿$$
$$g(x) = 1 - \frac{x}{¿} \quad (1)$$

El resultado de la segmentación obtenida por el autómata celular basado en el GrowCut es altamente dependiente de las condiciones y contenido de la imagen, así como de las semillas seleccionadas por el usuario, (He et al., 2014). Las imágenes satelitales poseen una gran cantidad de OG, lo que introduciría una dificultad importante en la colocación interactiva de las semillas. Por ello, se propone un algoritmo que permite seleccionar las semillas de manera automática.

**Selección de las Semillas**

Los OG corresponden a regiones de la imagen cuyos píxeles son conectados por presentar atributos similares. La frontera entre dos regiones se consigue al detectar alguna variación representativa entre alguno de los descriptores utilizados para su caracterización. Cada región presenta píxeles con descriptores homogéneos con cierta variación en su periferia por la mezcla con píxeles de regiones vecinas. Mediante el filtro de gradiente morfológico multiescala (Wang, 1997) aplicado en cada una de las bandas se obtiene una imagen sintética con valores bajos en las zonas homogéneas y valores más altos en las fronteras de las regiones. Las semillas representativas de cada OG se obtienen mediante el filtro morfológico de mínimos regionales del gradiente (Soille, 2004). La Fig. 2a presenta un detalle del gradiente, la Fig. 2b presenta la localización de las semillas.

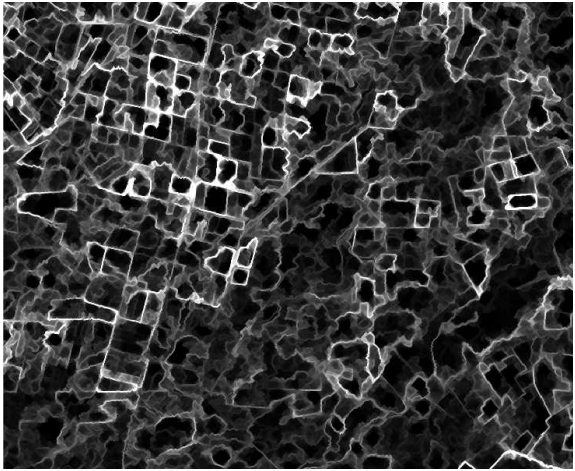
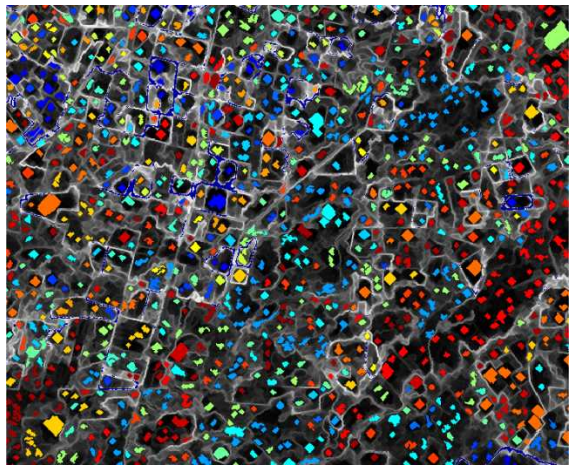

**Figura 2a**. Gradiente multiescala de la imagen. **2b**. Disposición de las semillas coloreadas según su etiqueta.

**Etiquetado de las Celdas Semilla**

En el estado inicial el autómata parte de las celdas semillas, a las cuales se les dota de una etiqueta y una fortaleza de ataque. Las semillas tendrán el máximo de fortaleza, es decir que no cambiarán de estado durante la evolución del autómata y deberán competir en la conquista del espacio de la imagen con las otras celdas semillas siguiendo la regla de evolución antes descrita.

Para el etiquetado de las celdas semillas se sigue el siguiente procedimiento:

• Cada grupo de píxeles semilla conectados estará representado por una firma espectral determinada por la moda en cada una de las bandas. La firma espectral obtenida será el descriptor de la semilla.

• Se calcula el índice de vegetación de la semilla y se incorpora como un descriptor suplementario. En las imágenes satelitales este parámetro contribuye con la discriminación más eficiente de los OG.

• Para el etiquetado de las semillas, los descriptores se agrupan en un número representativo de manera que los que tengan mayor similitud asuman una misma etiqueta. Después de varios ensayos se decidió tomar 20 grupos para reducir la sobre-segmentación de la imagen procesada. Se utilizó como técnica de agrupamiento el algoritmo k-means. Se hacen varias iteraciones tomando aleatoriamente vectores de inicio diferente para evitar los mínimos locales, entre los resultados obtenidos se toma aquel agrupamiento con mayor distancia entre los grupos formados.

• Solo las celdas semillas toman la etiqueta del grupo al cual fue asignada, mientras que el resto de las

celdas del autómata no tiene asignada ninguna etiqueta. En la Fig. 2b se muestra la disposición de las semillas sobre el gradiente de la imagen con un color distinto para cada etiqueta.
• Las celdas semilla toman el valor máximo de fortaleza, el resto de las celdas toma fortaleza nula.

La regla de evolución describe un procedimiento competitivo de invasión. Cada celda *p* es atacada por sus vecinas con una fuerza de ataque inversamente proporcional a la distancia entre sus descriptores $\bar{C}$. Si la fuerza de ataque supera su fortaleza, la celda p toma la etiqueta de la celda vecina q que la venció y adquiere como nueva fortaleza la fuerza de ataque $FA_q$. En cada iteración, cada celda tratará a su vez de invadir a las celdas vecinas y actualizar sus etiquetas. La evolución continúa mientras ocurran nuevas invasiones, cuando estas cesan se alcanza estabilidad y se obtiene la imagen segmentada.

**RESULTADOS**
El autómata evoluciona mientras ocurran cambios en el estado de alguna de las celdas, una vez alcanzada la estabilidad se tiene la imagen segmentada que se muestra en la Fig. 3. En este nivel de resultados, en la imagen etiquetada solo se agrupan píxeles que poseen atributos similares, condición que facilita un proceso ulterior de clasificación.

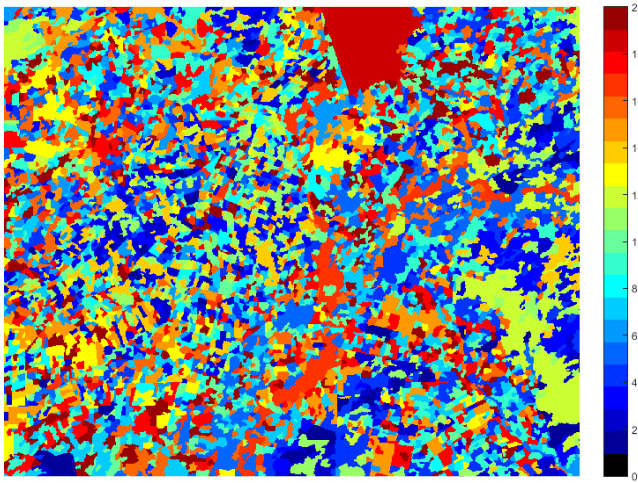

**Figura 3**. Imagen segmentada

En este caso se limitó a 20 el número de etiquetas, se considera que esta cantidad es suficiente para obtener un agrupamiento de celdas adecuado que evite la sobre-segmentación. En la Fig. 4a se muestra el detalle de la imagen segmentada y en la Fig. 4b los contornos de los segmentos superpuestos a la imagen en falso color infrarrojo, aquí se comprueba el ajuste de los contornos de los OG obtenidos

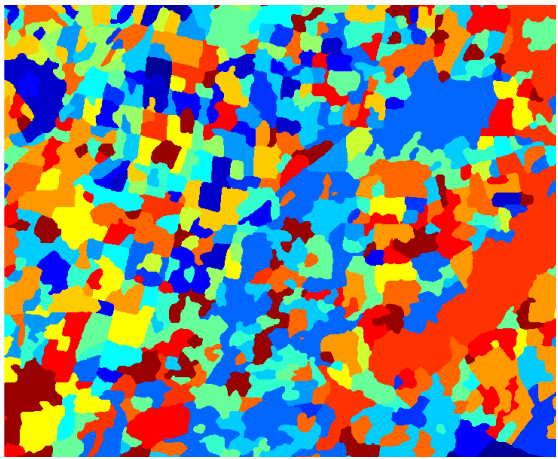
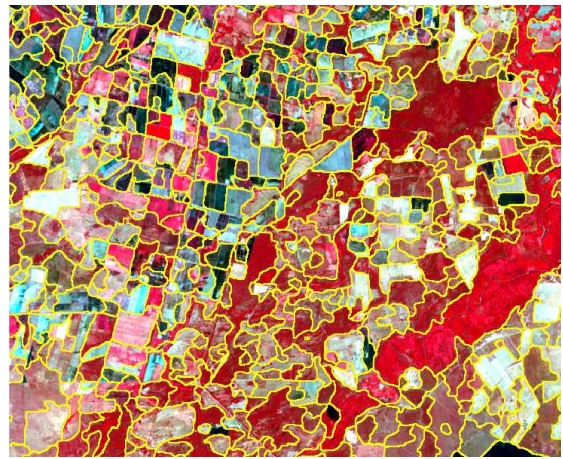

**Figura 4a**. Detalle de la imagen segmentada.    **4b.** Superposición de los contornos de los segmentos a la imagen en falso color infrarrojo

Se muestra en las imágenes de la Fig. 5 el detalle en otras regiones de la imagen con segmentos con características heterogéneas. En esta etapa del trabajo, el geógrafo especialista en interpretación visual de imágenes satelitales procedió a verificar el ajuste de los contornos de los OG en función de la escala de trabajo permitida por la resolución de la imagen.

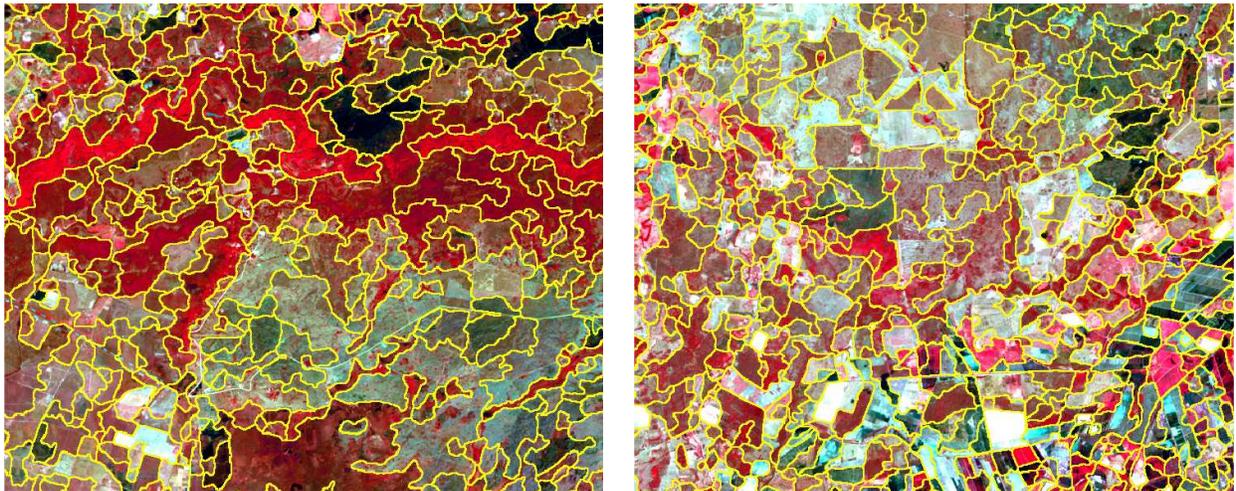

**Figura 5**. Detalle de la segmentación obtenida en diversas regiones de la imagen

## CONCLUSIONES

En este trabajo se presenta la utilización de los autómatas celulares en la segmentación automática de imágenes multiespectrales, los resultados obtenidos en imágenes donde se presentan regiones con características topográficas diversas detectan sus contornos de manera adecuada. Estos resultados contribuyen de manera importante en la reducción del tiempo que dedica el especialista durante la etapa preliminar de la clasificación orientada a objetos, que consiste en la revisión y edición de la segmentación obtenida por los programas que la ofrecen. Adicionalmente, respecto a la interpretación visual se consigue reducir la subjetividad introducida por parte del usuario. Quedan algunos ejes abiertos de investigación, como por ejemplo estudiar nuevas reglas de evolución para el autómata e introducir criterios relacionados con la escala de trabajo durante el etiquetado y la segmentación. La evaluación de la calidad de la segmentación solo se hace de forma cualitativa, ya que el flujo de trabajo considera la integración de la clasificación supervisada al proceso de segmentación para así evaluar sus resultados.

## REFERENCIAS